\documentclass{pasa}%
\pdfoutput=1
\usepackage{caption}
\usepackage{subcaption}
\usepackage{graphicx}
\usepackage{txfonts}
\usepackage{bm}
\usepackage[export]{adjustbox}

\renewcommand{\vec}[1]{\textsf{\textbf{#1}}}

\newcommand{\matrixtt}[4]{\left( \begin{array}{cc}#1&#2\\#3&#4\\\end{array} \right)}
\newcommand{\vectt}[2]{\left( \begin{array}{cc}#1\\#2\\\end{array} \right)}
\newcommand{\herm}{H}
\newcommand{\jones}[2]{\vec{#1}_{#2}}

\newcommand{\coh}[2]{\mathsf{{#1}}_{{#2}}}

\def\CSIRO{$^{1}$}
\def\Curtin{$^{2}$}
\def\CAASTRO{$^{3}$}
\def\CSIROPERTH{$^{4}$}
\def\ONSALA{$^{5}$}

\usepackage{amssymb,amsmath}
\title[MWA Tied-Array Processing]{MWA Tied-Array Processing I: Calibration and Beamformation}

\author[Ord S. M.  et al.]{
S.~M.~Ord\CSIRO$^,$\Curtin,
S.~E.~Tremblay\Curtin$^,$\CAASTRO,
S.~J.~McSweeney\Curtin$^,$\CAASTRO,
N.~D.~R.~Bhat\Curtin$^,$\CAASTRO,
C.~Sobey\Curtin$^,$\CSIROPERTH,
D.~A.~Mitchell\CSIRO$^,$\CAASTRO,
P.~J.~Hancock\Curtin$^,$\CAASTRO,
F.~Kirsten\ONSALA$^,$\Curtin
\\
\\
$^{1}$CSIRO Astronomy and Space Science, PO Box 76, Epping, NSW 1710, Australia\\
$^{2}$International Centre for Radio Astronomy Research (ICRAR),  GPO Box U1987, Perth, WA 6845, Australia\\
$^{3}$ARC Centre of Excellence for All-sky Astrophysics (CAASTRO)\\
$^{4}$CSIRO Astronomy and Space Science, PO Box 1130 Bentley, WA 6102, Australia\\
$^{5}$Department of Space, Earth and Environment, Chalmers University of Technology, Onsala Space Observatory, 439 92, Onsala, Sweden
}

\jid{PASA}
\doi{10.1017/pas.\the\year.xxx}
\jyear{\the\year}

\usepackage{fancyhdr, setspace, color, soul}
\usepackage{graphicx}
\usepackage{amsmath}
\usepackage[notes,backend=biber,maxbibnames=99]{biblatex-chicago}
\DeclareBibliographyAlias{webpage}{online} 

\AtEveryBibitem{\clearlist{language}}
\AtEveryBibitem{%
   \ifentrytype{online}
    {}
   {\clearfield{urlyear}\clearfield{urlmonth}
        \clearfield{doi}          
        \clearfield{note}
        \clearfield{urlday}}}

\begin{document}%
\begin{abstract}
The Murchison Widefield Array is a low-frequency Square Kilometre Array precursor located at the Murchison Radio--astronomy Observatory in Western Australia. Primarily designed as an imaging telescope, but with a flexible signal path, the capabilities of this telescope have recently been extended to include off-line incoherent and tied-array beam formation using recorded antenna voltages. This has provided the capability for high-time and frequency resolution observations, including a pulsar science program. This paper describes the algorithms and pipeline that we have developed to form the tied array beam products from the summation of calibrated signals of the antenna elements, and presents example polarimetric profiles for PSRs J0437$-$4715 and J1900$-$2600 at 185\,MHz.
\end{abstract}
\begin{keywords}
instrumentation: interferometers, techniques: interferometric, pulsars: general, pulsars: individual (PSR J0437--4715, PSR J1900--2600)
\end{keywords}
\maketitle%

\section{Introduction}

The Murchison Widefield Array Phase 1 (MWA; Tingay et al. 2013)\nocite{tingay:2013}, consists of 2048 dual--polarisation dipole antennas arranged into 128 aperture array `tiles' to form a connected element interferometer.  It operates between approximately 80 and 300\, MHz and was primarily designed as an imaging telescope. The array has now been upgraded to Phase 2, being described in Wayth et al (2018)\nocite{wayth:2018}, which deploys 256 tiles, but with only 128 available at any given time. Phase 2 supports two configurations, one a compact configuration with baselines in general shorter than in Phase 1, and an extended configuration, with baselines substantially longer than in Phase 1. We present the pipeline and algorithms employed to provide array voltage beams using 128 antennas, for high time resolution polarimetric observations of pulsars. This pipeline was originally developed for MWA Phase 1, but is also used in Phase 2 operations.

As an interferometer, the signals from all of the 128 dual-polarisation antennas are brought together in the MWA correlator (Ord et al. 2015)\nocite{ord+15}, and the {\em cross-power spectrum} is measured.  The output from the correlator is accumulated for an integration period of at least 0.25 seconds and usually longer. These visibility sets are used to form images of the radio sky. The time resolution of imaging interferometers is typically too coarse for pulsar observations, and the full cross power spectrum is a very large data set for telescopes with a large number of elements. Full correlation of radio arrays is also computationally challenging and early observations of pulsars using radio telescope arrays were performed by summing the elements into phased-array beams; indeed this is how they were serendipitously discovered (Hewish et al. 1968\nocite{hew+68}). As the pulsar field matured, observations were preferentially made at higher frequencies with single dishes. Recently, interest in the Epoch of Re-ionisation has driven the design and production of next--generation low--frequency radio arrays with extensive collecting area and full correlation between elements (e.g. the MWA and LOFAR; van Haarlem et al. 2013\nocite{LOFAR}; and SKA-LOW\footnote{ As detailed at http://www.ska.gov.uk} in the design phase). Pulsar observations are key science cases for LOFAR and the SKA, and in this paper we outline the beamforming capability that has been developed for the MWA. 

For the beamformer the signals from each antenna are recorded by the MWA Voltage Capture System (VCS; Tremblay et al 2015\nocite{tre+15}). The VCS system records channelised data -- 3072 channels across 30.72~MHz of bandwidth, for each of the 256 inputs. After the signal streams are recorded, they are calibrated and combined into a single, channelised, dual-polarisation voltage, tied-array, {\em pencil beam}. In beamforming we are attempting to measure the sky brightness in a single look direction. This does not require that we form the full cross correlation between all elements; we need only form a sum over all the elements, producing a data product that is similar to the single pixel of an interferometric image. This is a much less compute intensive operation than all-to-all correlation. However, the sum has to be performed {\em coherently}, and separately for each look-direction, otherwise the contributions from each antenna will not add constructively. This requires precise instrumental calibration to ensure minimum loss of sensitivity. 
 
Initial pulsar observations made with the MWA did not utilise beamforming. It is much simpler to avoid precise calibration and coherent addition and to detect the signal from each antenna and sum this detected product. This is called the {\em incoherent-beam} and has been used to perform a census of southern hemisphere pulsars (Xue et al 2017\nocite{xue:2017}) and to observe the properties of individual pulsars and the intervening interstellar medium (Bhat et al 2014\nocite{bhat:2014}). The disadvantage of this method is that each antenna has such low intrinsic gain that the incoherent sum has a large noise component, resulting in comparatively low sensitivity.  The advantage of the incoherent sum is the wide field of view, which may contain many objects simultaneously, and its computational simplicity. We examine the relative performance of the incoherent and coherent sum for the MWA in \S \ref{sec:SNR}.

\begin{figure}[htbp] 
   \centering
   \includegraphics[width=3in ]{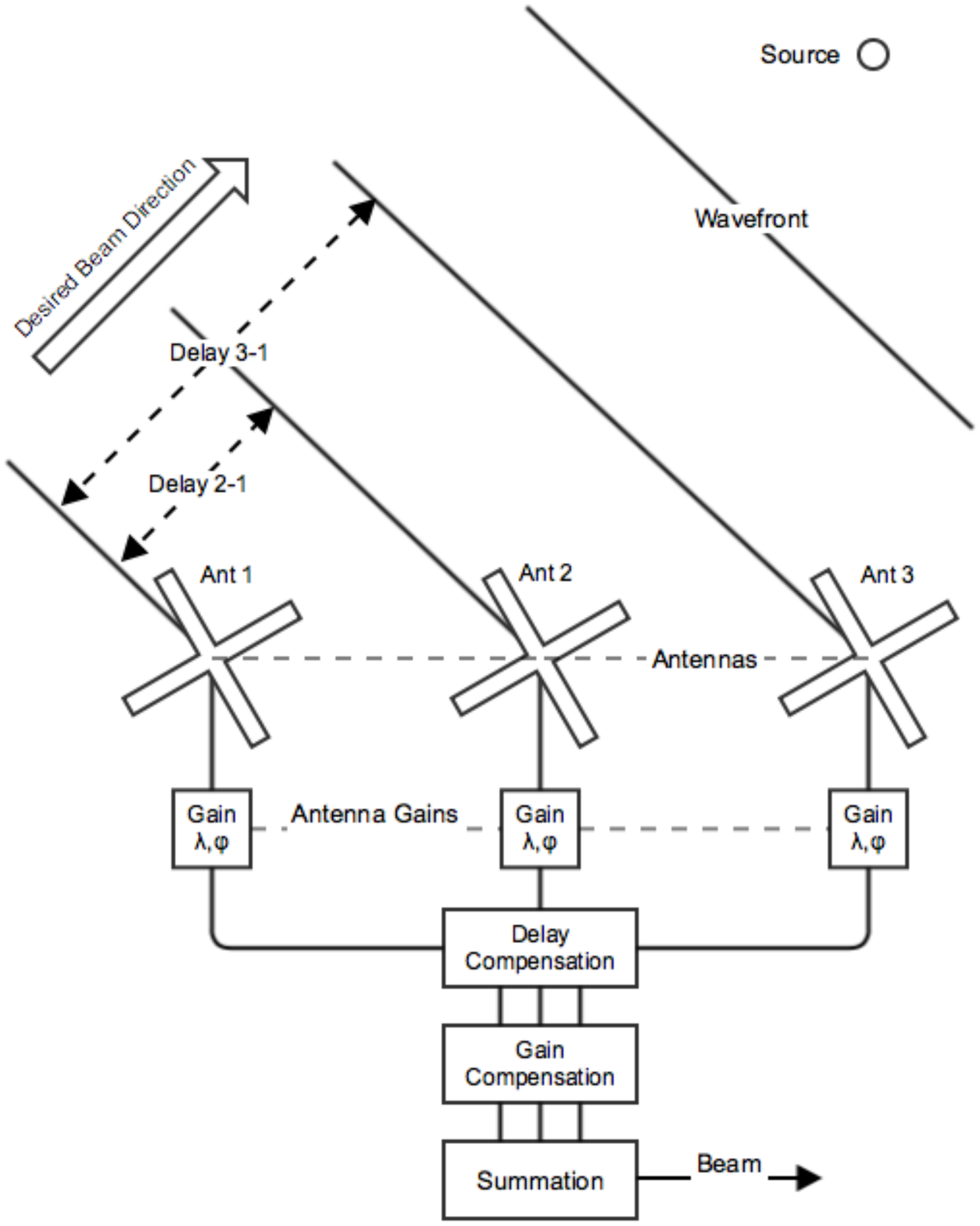} 
   \caption{A schematic of a 3--element, dual-polarisation, beamformer. In order to coherently sum the 6 signals from the 3 antennas, one must correct for the geometric delays, the antenna gains and the cable delays. Both the cable and geometric delays can be measured or calculated, but the gain is unknown and must be determined by the calibration process.}
   \label{fig:1}
\end{figure}

In order to direct the tied-array beam to a given look direction the outputs of all antennas must be combined, incorporating a compensating delay for {\em beam steering}, cable length differences, and the phase of the antenna complex gain (see Figure \ref{fig:1} for a schematic of a 3--element beamformer). In general the delay is not an integer number of discrete samples, so is often applied as both a whole sample delay correction and a residual phase correction. The calibration of the tied array beam utilises the {\em Real Time System} (RTS: Mitchell et al 2008\nocite{mitchell:2008}) in order to obtain estimates of the complex gain of each constituent antenna. 

Following the delay and calibration compensation the individual antenna voltage streams are combined into a single dual-polarisation beam, which can then be packed into two file formats. Either an undetected dual-polarisation format, VDIF (Whitney et al 2009\nocite{whitney:2009}). Or the time series can be detected, formed into Stokes parameters and packed into PSRFITS (Hotan et al 2004) format. All subsequent analysis is performed by the \textsc{PRESTO} (Ransom 2011\nocite{ransom:2011}), \textsc{DSPSR} (van Straten and Bailes 2010\nocite{vSB:2010}) and \textsc{PSRCHIVE} (Hotan et al. 2004\nocite{hotan:2004}) software packages.

The format of this paper is as follows. We first discuss the practical determination and application of the delay corrections. Then we outline the calibration process, which determines the antenna Jones matrices. We then describe the application of the compensating delays and the calibration, and the formation of the beams. We finally present full polarimetric profiles of pulsars obtained using the MWA tied-array beamformer in both coherent and incoherent mode and discuss their properties.

\section{Data Acquisition, Array Calibration and Beam Formation}

Calibration is the act of determining the relative {\em gain} of each constituent antenna. In this context it is performed for the purpose of maintaining coherence in the beamforming process. The discussion of array calibration in this paper follows the Jones matrix formalism popularised for radio astronomy in the series of papers by Hamaker, Sault and Bregman (1996, 1996a, 1996b and 2000, henceforth HSB) \nocite{hamaker:1996}\nocite{sault:1996}\nocite{hamaker:1996b}\nocite{hamaker:2000} and revisited by Smirnov (2011a, 2011b, 2011c)\nocite{smirnov:2011a}\nocite{smirnov:2011b}\nocite{smirnov:2011c}. We will repeat the salient points of this description as it provides useful background, but direct the reader to these papers for a more complete treatment. Firstly, we outline the process by which the data streams are acquired and the delay model constructed.

\subsection{Data Acquisition}

The subsystem that records the baseband data, the so-called Voltage Capture System (VCS), and the data produced by the system are described in detail in Tremblay (2015) \nocite{tre+15} and will be outlined here on a cursory level. 

The VCS splits off the complex data after they have been channelised into 10\,kHz and 100\,$\mu$s samples by a polyphase filterbank. These data pass through a set of 16 servers, each of which possesses 2 x 1.44\,TB RAIDs (Redundant Arrays of Independent Disks). The channelised data can be written out to these RAIDs in real-time for $\sim 100$ minutes before reaching their capacity limit. The data are then moved into the MWA data archive hosted by the Pawsey Supercomputing Centre in Perth, Australia. 

VCS files each consist of 4-bit + 4-bit complex voltages from every antenna, polarisation, 10\,kHz channel, and 100\,$\mu$s; each 242 MB in size. 32 of these files are recorded for each second of MWA-VCS observations, generating a data volume of $\sim$28\,TB/hour.


\subsection{Initial Delay Correction}
\label{sec:delay}

We implement all of the features of delay tracking and beam formation as a software operation on the recorded telescope voltages. The metadata for an MWA observation are held in the MWA observing database and can be accessed in the form of a FITS table. For the purposes of the delay calculations the data retrieved from this file are; the locations of the antennas, the cable lengths, and the pointing of the telescope. The analysis is simplified by considering a reference position to which all signals will be delayed, then each antenna delay-distance can be considered a baseline, and measured in the same way as the typical interferometric ($u, v, w$) coordinates. The geometric delay-distance, measured in metres, for a given antenna, ${j}$, is simply the $w$ coordinate as calculated by:

\begin{align}
w_j &= \cos{(\delta)}\cos{(H)}{X_j}  - \cos{(\delta)}\sin{(H)}{Y_j} + \sin{(\delta)}{Z_j}, 
\label{eq:w}
\end{align}

\noindent where $({X_j},{Y_j},{Z_j})$ is the $j^{\mathrm{th}}$ antenna's position in {\em local} geocentric coordinates. The local frame is where $Z$ points north, $X$ points through the equator from the geocentre along the local meridian and  $Y$ is east. This is a geocentric, earth fixed coordinate system, except the $X$ axis is pointing toward the local meridian and not Greenwich. $H$ and $\delta$ are the hour angle and declination of the look direction respectively. 
The time delay, with units of seconds, including the electrical cable length ($L_j$) is:
 
 \begin{equation}
 \Delta{t}_{j} =  \left( w_{j} + L_{j} \right) / c\,,
 \label{eq:t}
 \end{equation}
 
 \noindent where $c$ is the speed of light. For all the MWA antenna positions and cable lengths, and with the time resolution of the captured data (100\,$\mu\mathrm{s}/\mathrm{sample}$), all of the antenna data streams arrive within a sample-time. No whole sample delays are required. It is computationally simpler to express this time delay as a phase shift. To do this firstly we need to consider time measured in {\em samples}, rather than seconds. We then need to consider the frequency of the observation, as the phase shift is a function of frequency. In our case, the raw data captured by the VCS system is already channelised, so this correction can be applied per channel. In practice, with time measured in seconds and frequency in Hz all unit conversions cancel for the phase calculation. The phase correction for a channel, $n$ with centre frequency $f_{n}$ of antenna ${j}$ required to compensate for beam steering and cable delays is:

\begin{equation}
\phi_{{j},{n}} = 2\pi \Delta{t_{j}} {f_n} \,\,\, \mathrm{rad}
\label{eq:phi}
\end{equation}
\subsection{Fringe Rate}
\label{sec:fr}
As the Earth rotates, the projection of each baseline changes with respect to the phase tracking centre. For a phase centre at fixed celestial coordinates, the only time-variable quantity in Equation \ref{eq:w} is the hour angle $H$. Differentiating Equation \ref{eq:w} with respect to time:

 \begin{equation}
 \frac{dw_j}{dt} = -1 \times \left(\sin{(H)}{X_j} + \cos{(H)}{Y_j}\right)\cos{(\delta)}\frac{dH}{dt},
 \end{equation}

\noindent and the only time-variable quantity in Equation \ref{eq:phi} is the $w_j$ in $\Delta{t_{j}}$. The fringe rate is therefore:

\begin{align}
\frac{d\phi_{{j},{n}}}{dt} &= -2\pi\left(\sin{(H)}{X_j} + \cos{(H)}{Y_j}\right)\cos{(\delta)}\frac{dH}{dt}\frac{f_n}{c}, \\
\frac{dH}{dt} &=  \frac{2\pi}{8.64\times10^4}.
\end{align}

\noindent For an east-west (E-W) baseline, the fringe rate simplifies to:

\begin{align}
\frac{d\phi_{{j},{n}}}{dt} &= -4.6\times10^{-4}{Y_j} \cos{(H)}\cos{(\delta)}\frac{f_n}{c}.
\end{align}

We assign a reference position, to which all antennas will be delayed, to be near the centre of the array. The maximum ``baseline'' therefore, in the specific case of the MWA Phase 1, is approximately 1.5\,km.  Given a  maximum frequency of approximately 300\,MHz, the maximum fringe rate (with source on the celestial equator, crossing the local meridian) is approximately 0.5 rad\,$\mathrm{s}^{-1}$. The MWA Phase 1 antenna distribution is centrally condensed with most antennas being within a few hundred metres; which implies that for the vast majority of baselines the rate is much less than this. The MWA Phase 2 offers an extended layout allowing baselines as long as 5\,km. This will approximately double the maximum fringe rate for these antennas to $\sim$1 rad\,$\mathrm{s.}^{-1}$

\subsection{Gain Calibration}
\label{sec:gaincal}

Each antenna in the array has a complex gain, imparting a phase turn on the incoming voltage stream. This serves to decorrelate the sum of the antenna signals. The gain calibration process is an attempt to determine this instrumental response. Since the work of HSB it is commonplace to describe the instrumental Jones matrix of antenna $j$ as a combination of antenna-based Jones matrices:

\begin{equation}
\jones{J}{j} = \jones{G}{j}\jones{D}{j}\jones{A}{j}\jones{P}{j}\jones{T}{j}\jones{I}{j}\jones{F}{j},
\label{eq:jones}
\end{equation}

\begin{itemize}
 \item $\jones{G}{}$ -- {\em electrical gains}: Usually diagonal; can be direction dependent. 
 \item $\jones{D}{}$ -- {\em leakage}: how much of one polarisation is detected in the orthogonal polarisation; direction dependent. 
 \item $\jones{A}{}$ -- {\em antenna effects}: any peculiarities of the antenna that are a-priori known.  
  \item $\jones{P}{}$ --  {\em parallactic angle}: rotation of the feed with respect to the sky: rotation matrix, time variable.
 \item $\jones{T}{}$ --  {\em tropospheric gain}: diagonal and polarisation independent, time variable.
 \item $\jones{I}{}$ --  {\em ionospheric gain}: ionospheric opacity, time variable. 
 \item $\jones{F}{}$ -- {\em Faraday rotation}: due to the ionosphere: rotation matrix; direction dependent, time variable.
\end{itemize}%

In order to compensate for the different antenna gains, they must be calibrated; that is, placed on the same relative, or absolute, amplitude and phase scale. The individual antennas that comprise the MWA are of low intrinsic gain; the sky at low frequencies is dominated by bright diffuse synchrotron radiation, and the antennas do not have calibrated noise diodes to be used as model sources. The antennas cannot therefore be calibrated individually, and must be calibrated as an interferometer. This allows baselines to be combined to increase signal to noise (shorter baselines to be excluded from the sum to reduce the contribution from diffuse structure) and a component model of the sky brightness can be used as the model in the calibration process. However, in order to calibrate the MWA using this method the raw voltages, as captured for beamforming, must be correlated to form visibilities from which the calibration solution can be obtained. The calibration is obtained either via a short calibration scan performed on a nearby calibrator field, or an in-field calibration is used. The FoV of the MWA is very large and a given field contains many sources suitable for calibration (see \S \ref{sec:cml}). A salient feature of the MWA is that it employs a software correlator, which can be run offline on the recorded data, generating exactly the same products as are generated on-line (Ord et al. 2015).\nocite{ord+15} In the case where dedicated calibration observations are used, experience has shown that calibration scans with the sources close to transit are generally better, generating higher signal-to-noise ratio tied-array beams (Xue et al submitted). Therefore often a single high quality calibration observation will be made and will serve for all observations during a session, typically a few hours in duration. 

Following HSB and Smirnov (2011a,b,c), for convenience we will collapse Equation \ref{eq:jones} and write the instantaneous response of an antenna, $j$, to an incident electric field vector, $\vec{e} =( e_{x},e_{y})^H$, where $H$ is the Hermitian transpose, as the antenna voltage $\vec{v}_{j}$:

\begin{equation}
\vec{v}_{j} = \jones{J}{j}\vec{e}.
\label{eq:single_jones}
\end{equation}

The action of correlating the voltages from any two antennas ($j$ and $k$), is performed as an outer product between the antenna pairs, performed for all unique pairs, and in each frequency channel. a process which produces the visibility matrix:  

\begin{equation}
\coh{V}{jk}  = \langle \vec v_j \vec v^\herm_k \rangle =  \jones{J}{j}\langle \vec{e}\vec{e}^H \rangle \jones{J}{k}^H.
\label{eq:RIME}
\end{equation}

\noindent Where $\langle . \rangle$ denotes an average over some time and frequency interval the Jones matrices can be considered constant. The outer-product $\langle \vec{e}\vec{e}^H\rangle$ results in the {\em coherency matrix} which for a single look-direction has a direct relationship to the Stokes parameters given by:

 \begin{equation}\label{eq:IQUV}
 \langle \vec{e}\vec{e}^H\rangle =    \matrixtt{\langle e_x e^*_x\rangle }{\langle e_x e^*_y\rangle }{\langle e_y e^*_x\rangle }{\langle e_y e^*_y\rangle }
    = \frac{1}{2} \matrixtt{I+Q}{U+iV}{U-iV}{I-Q}.
\end{equation}

\noindent Using Equations \ref{eq:RIME} and \ref{eq:IQUV} we can relate the polarisation state of a calibrator source to the measured visibilities, via the Jones matrices of the antennas.



\subsubsection{Antenna Jones Matrix Determination and Calibration}
\label{sec:cml}

With this framework we will now outline the mechanisms by which the Real-Time-System (RTS) Calibration and Measurement Loop (CML) determines the unknown Jones matrices from the measured visibilities. The CML was specifically designed to incorporate direction dependence and ionospheric corrections into the gain determination. At MWA observing frequencies ionospheric phase shifts are considerable. Since these are variable in time and direction, they must be decoupled from instrumental phase shifts to limit the number of free calibration parameters. The CML assumes that the phases scale linearly with wavelength and, for a given frequency and calibrator, have a linear trend across the relatively small array. The CML can then track a single wavelength-squared-dependent refractive offset for each calibrator source, while building up antenna gain information on slower time scales. The CML starts by finding the brightest calibrators and estimating their apparent flux densities (in the form of coherency matrices) from a source catalogue. The phase at which a calibrator enters the visibility is equivalent to the geometric delay discussed in \S \ref{sec:delay}; it is calculated as the scalar product between the baseline vector and the look-direction, which includes a component due to catalogue position and the refractive ionospheric offset. Denoting the  combined phase shift for calibrator $a$, at the frequency $f_n$, as $\phi_{jk,a,f_{n}}$, we can form model visibilities as the sum over calibrators:

\begin{equation}
\coh{V}{jk, f_{n}}^{\mathrm{model}} = \sum_{a=1}^{N_{C}} \jones{J}{j,a}^{ }\jones{P}{jk,a}^{ } \jones{J}{k,a}^H \exp\{-i \phi_{jk,a,f_{n}}\}
\end{equation}

\noindent where $\jones{P}{jk,a}^{ } $ is the coherency matrix for calibrator $a$. We will use this simple summation over calibrator sources for clarity, but in practice we can extend the sky model over more than the $N_{C}$ calibrators, can include a tapering term in $uv$ to convolve a sky component with a known morphology, and can form a given calibrator from a summation over multiple neighbouring components. As the calibration components can be complex structures, the model is generally a function of baseline. The full-sky model is then subtracted from the visibility set.
\begin{eqnarray}
\coh{V}{jk, f_{n}}^{\mathrm{residual}} &=& \coh{V}{jk, f_{n}} - \coh{V}{jk, f_{n}}^{\mathrm{model}}.
\label{eqn:subtract}
\end{eqnarray}
The calibrator (or more complex model), $a$, is then added back in at its current best-fit position. To form a visibility set which is dominated by the calibrator.
\begin{eqnarray}
\coh{V}{jk,a, f_{n}} &=& \coh{V}{jk, f_{n}}^{\mathrm{residual}}  + \jones{J}{j,a}^{ }\jones{P}{jk,a}^{ } \jones{J}{k,a}^H \exp\{-i \phi_{jk,a,f_{n}}\} .
\label{eqn:add_back}
\end{eqnarray}
The visibility set is then rotated to the calibrator position and undergoes a small amount of averaging in time and frequency to increase the S/N.
\begin{eqnarray}
\coh{V}{jk,a, f_{n}}^{\prime} &=& \coh{V}{jk, f_{n}}^{\mathrm{residual}}\exp\{i \phi_{jk,a,f_{n}}\} + \jones{J}{j,a}^{ }\jones{P}{jk,a}^{ } \jones{J}{k,a}^H.
\label{eqn:rotate}
\end{eqnarray}

\noindent The measured $\coh{V}{jk,a, f_{n}}^{\prime}$ can be directly compared to a model of the measured visibility matrix for the source $a$. In other words, we are attempting to find a gain solution for which  $ \coh{V}{jk, f_{n}}^{\mathrm{residual}}\exp\{i \phi_{jk,a,f_{n}}\} = 0$ and consequently:

\begin{eqnarray}
\jones{J}{j,a}^{ }\jones{P}{jk,a}^{ } \jones{J}{k,a}^H = \coh{V}{jk, f_{n},a}^{\prime}. 
\label{eqn:problem}
\end{eqnarray}

This is an implementation of the {\em peeling} scheme introduced by Noordam et al 2004\nocite{noordam:2004} and is repeated iteratively for a list of $N_{c}$ calibrators. 
There are as many implementations of the solution to Equation \ref{eqn:problem} as there are astronomical software packages. Here we will present a derivation of the gain solution scheme used in the RTS. The scheme is essentially a fully polarimetric implementation of  {\em antsol}, as described in Bhatnagar and Nityananda (2001)\nocite{bhatnagar:2001} and references therein.  Equation \ref{eqn:problem}  can be linearised by assuming that $\jones{P}{jk,a}^{ }$ is known, which by definition is one of the components of the sky model, and all the $\jones{J}{k,a}^{n-1} $ are known from a starting estimate, or the previous iteration $(n - 1)$. We are solving for the error matrix $\jones{E}{j} $, that ideally corrects a previous iteration of the antenna gain - to a better estimate for the current iteration:
\begin{eqnarray}
\jones{J}{j,a}^{n} = \jones{E}{j}\jones{J}{j,a}^{n-1}
\end{eqnarray}
This system is overdetermined and can be solved in a {\em least squared} sense by minimising the objective function:  
\begin{eqnarray}
S(\jones{E}{j}) =  \sum^{N_{A}}_{jk,j \neq k} \|  \jones{E}{j}\jones{J}{j,a}^{}\jones{P}{jk,a}^{ } \jones{J}{k,a}^H - \coh{V}{jk,a, f_{n}}^{\prime} \|^{2},
\label{eqn:obj1}
\end{eqnarray}
 
\noindent where all $\jones{J}{}$ are from the starting estimate or previous iteration. As this is a full polarimetric treatment all the terms in Equation \ref{eqn:obj1}  are $2\, \times \,2$ matrices, therefore minimising this equation in order to find the best set of gain solutions, in a least squares sense, is a non-trivial operation. We have used the symbol for a norm $(\|.\|$) as to express the size of the squared difference between the model and the measurements, but not indicated which norm we should use, or how it should be calculated. In the RTS the chosen norm is the Frobenius, or Hilbert-Schmidt norm, $\|.\|_F$.

\begin{eqnarray}
\|\jones{A}{}\|_F = \sqrt{\mathrm{tr}(\jones{A}{}\jones{A}{}^{H})}
\end{eqnarray}
 
\noindent So the objective function becomes formally:
 
\begin{equation}
\begin{aligned}
S(\jones{E}{j}) &=&  \mathrm{tr} (\sum^{N_{A}}_{jk,j \neq k} (\jones{E}{j} \jones{J}{j,a}^{ }\jones{P}{jk,a}^{ } \jones{J}{k,a}^H - \coh{V}{jk, f_{n}}^{\prime} )\,\times \\ 
 & & (\jones{E}{j} \jones{J}{j,a}^{ }\jones{P}{jk,a}^{ } \jones{J}{k,a}^H - \coh{V}{jk, f_{n}}^{\prime})^{H} )
 \end{aligned}
 \end{equation}
 For clarity, we will denote the predicted, or {\em model} visibility $ \jones{J}{j,a}^{ }\jones{P}{jk,a}^{ } \jones{J}{k,a}^H$ as $\jones{M}{jk,a}$
 \begin{equation}
 \begin{aligned}
  S(\jones{E}{j}) =  \mathrm{tr}( \sum^{N_{A}}_{jk,j \neq k}(  \jones{E}{j}\jones{M}{jk,a}\jones{M}{jk,a}^H\jones{E}{j}^{H} -  \jones{E}{j}\jones{M}{jk,a}\coh{V}{jk,a, f_{n}}^{\prime H} \\ 
  - \coh{V}{jk,a, f_{n}}^{\prime} \jones{M}{jk,a}^H\jones{E}{j}^{H} + \coh{V}{jk,a, f_{n}}^{\prime} \coh{V}{jk,a, f_{n}}^{\prime H})) \\
 \label{eqn:obj2}
 \end{aligned}
 \end{equation}
 
The minimum of Equation \ref{eqn:obj1} is found by differentiating Equation \ref{eqn:obj2} with respect to the error term $\jones{E}{j}$ and equating the result to zero. In performing this differentiation three points need to be considered. Firstly, a complex variable and its conjugate are independent. Secondly, we are assuming all $\jones{E}{j}$ are independent. Finally, in considering the derivative of a $\mathrm{trace}$ that contains a matrix product the following holds:

\begin{equation}
\frac{\partial \, \mathrm{tr} (\jones{A}{}\jones{B}{})}{\partial \jones{B}{}} = \jones{A}{}^{H}
\end{equation}

\begin{equation}
\begin{aligned}
\frac{\partial{S(\jones{E}{j})}}{\partial{\jones{E}{j}}} = \sum^{N_{A}}_{k, k \neq j}\left[\jones{E}{j}\jones{M}{jk,a}\jones{M}{jk,a}^{H} - \coh{V}{jk,a, f_{n}}^{\prime}\jones{M}{jk,a}^{H}\right]. 
\end{aligned}
\end{equation}
Setting this derivative to zero reveals:
\begin{equation}
\begin{aligned}
\sum^{N_{A}}_{k, k \neq j}\left[\jones{E}{j}\jones{M}{jk,a}\jones{M}{jk,a}^{H} \right] = \sum^{N_{A}}_{k, k \neq j}\left[\coh{V}{jk,a, f_{n}}^{\prime}\jones{M}{jk,a}^{H}\right]. 
\label{eqn:obj3}
\end{aligned}
\end{equation}
As the summation on the L.H.S. of Equation \ref{eqn:obj3} is over $k$ only, we can move the $\jones{E}{j}$ outside the summation to obtain:
\begin{eqnarray}
\jones{E}{j} = \sum^{N_{A}}_{k, k \neq j}\left[\coh{V}{jk,a, f_{n}}^{\prime}\jones{M}{jk,a}^{H}\right] \times \left[\sum^{N_{A}}_{k, k \neq j}\jones{M}{jk,a}^{ }\jones{M}{jk,a}^{H} \right]^{-1}
\label{eqn:soln}
\end{eqnarray}

\noindent This shows that the update matrix for antenna $j$ is formed by the sum of all visibilities containing that antenna, weighted by the model, divided by the square of the weights. Therefore in the case where the model matches the visibilities the update matrix is the identity matrix. With this scheme we start with an estimate for all antenna gains $\jones{J}{j,a}$, then via Equation \ref{eqn:soln} obtain a new estimate of $\jones{E}{j}$. We then use this to update our estimate of $\jones{J}{j,a}$. 

There are multiple cycles of the calibration and measurement loop, we iterate over each gain calibrator. Full gain calibration as described above is only applied to a small number of calibrators, but many can be used to measure ionospheric offsets. In the beamforming implementation, where we only require a good gain estimate in a single direction for the purpose of beamforming the dominant ionospheric shift is accounted for by the bulk offset of the brightest calibrator from its catalogue position. This shift is generally incorporated into the gain solutions. 
Although in this implementation we only require the solution in a single direction, we benefit from employing multiple calibrators in two ways: firstly in the scheme where we treat each calibrator separately we achieve a better gain solution by more accurately removing the contribution from the contaminating sources. Secondly and more commonly, we use multiple calibrators as components in a field based calibration scheme where the model contributions are summed into a {\em single} model before fitting. This has the benefit of increasing the SNR of the model as it includes more flux. The disadvantage is that any errors in the model components are not individually fit for. In the case of multiple calibrators we generally decouple the solutions by not simultaneously calibrating on and subtracting sources, except where complete subtraction of contaminating sources in warranted, as there only so far you can go with this scheme before you start removing too many degrees of freedom from the dataset. There are more sophisticated schemes that can simultaneously solve for multiple directions (SAGEcal as described in Kazemi et al 2011 and Yatawatta et al 2009. In tests we have found that we converge to the same solutions as these more sophisticated schemes, but generally we require more iterations. Our iterations are however cheaper in terms of operations.

\subsubsection{Update Rates and Degeneracies}

In practice we average the previous iteration with the new prediction to obtain the next estimate.
\begin{eqnarray}
\jones{J}{k,a}^{n} = ((1-\alpha) + \alpha\jones{E}{j})\jones{J}{j,a}^{ n - 1},
\end{eqnarray}
This update has been presented by previous authors: HBS, Mitchell et al (2008), and discussed in detail by Salvini and Wijnholds (2014)\nocite{salvini:2014}.
\noindent we have found that this scheme is very stable, provided $\alpha$ is less than 0.5. In the case of low S/N we have found that reducing the update rate is useful in maintaining convergence.

The whole scheme is predicated on the relationship between the calibrator coherency matrix and the visibilities, Equation \ref{eqn:problem}. However it is possible to replace the calibrator $P$, via a similarity transform of the following form:
\begin{eqnarray}
\jones{U}{}\jones{P}{}\jones{U}{}^H = \jones{P}{}
\label{eqn:similar}
\end{eqnarray}
This is possible for only a small set of $\jones{U}{}$ and, or, $\jones{P}{}$. For  example if $\jones{U}{}$ is a rotation and $\jones{P}{}$ is unpolarised then Equation \ref{eqn:similar} clearly holds. This amounts to saying if the calibrator is unpolarised then there is an unconstrained rotation that can be absorbed into the calibration solutions. Note that this is true even if there is a significant amount of instrumental polarisation as it is the diagonal nature of the model that permits the degeneracy. There is another degeneracy, as the calibration is also insensitive the row exchange operator providing the calibrator is unpolarised. Generally, the second case is easier to spot as this increases the value of the off-diagonal terms in the calibration, this can manifest as linear polarisation suddenly transferring between $Q$ and $U$. We cannot alleviate the first degeneracy without a source of known polarisation, but we can reduce the likelihood of the second by using starting estimates that are close to the actual solutions via a reasonable model of the antenna.

\subsection{Beam Formation}

The raw voltages for each coarse, 1.28~MHz, channel are retrieved from storage and the formation of the tied array beam proceeds as a pipeline operation after a calibration solution is obtained. Each coarse channel is beamformed independently. A gain solution is obtained for each antenna using the geometric delay to the pointing centre and the calibration solution; every sample from every antenna is multiplied by its gain solution to equalise the antenna gains and phase all antennas to the same direction. Either the Stokes parameters are formed, or the single dual--polarisation tied-array voltage beam is written to disk. The retrieval of the raw data is an implementation-dependent task and can be arranged via discussion with the MWA project (contacted via http://mwatelescope.org). In this section of the paper we will discuss the determination and application of the gain solution and the formation of both the Stokes parameters and the tied-array voltage beam.

\subsubsection{Delay Compensation}

Firstly, the process outlined in \S \ref{sec:delay} is employed: given the arrival time of the voltage sample, and the array layout, the ($u, v ,w$) for each antenna relative to a reference antenna are determined.  The $w$ being that used in Equations \ref{eq:w} and \ref{eq:t} to determine the geometric delay for each antenna. The relative cable length between the antenna in question and the reference antenna is obtained and associated with the {\em electrical length} of this cable difference, this is then turned into a time delay. Both these time delays are combined and converted into a phase shift using Equation \ref{eq:phi}. This phase correction is determined for each 10~kHz channel and updated every second to track the source.

As has been noted, in \S \ref{sec:fr}.  the geometric delay is changing at a rate of a 0.5 rad\,$\mathrm{s}^{-1}$ on the longest baseline at the highest frequency, but is typically much slower due to the centrally condensed array layout. As we are only updating the delay compensation every second, the phased array beam will decorrelate to some degree.  To minimise this we calculate delays at the centre of each second, which limits the phase error to a maximum of 0.25 rad, in the worst case. This is an acceptable level of phase error: quarter of a radian of phase error corresponds to a 1\% drop in signal to noise ratio. The extended configurations may require an increase to this update rate, which can be accommodated.

\subsubsection{Gain Compensation}

The next step is to incorporate the antenna gains into the delay model. The MWA beamformer can use antenna gain calibrations obtained via a variety of methods. We predominately use solutions obtained by the RTS, as introduced in \S \ref{sec:gaincal}, but can also read the calibration information determined by the {\em calibrate} tool (Offringa et al 2014) \nocite{off:2014} used by the GLEAM survey team (Hurley-Walker et al 2017)\nocite{hw:2017}. These two methods incorporate a level of direction dependence in that they account for the MWA beam model and the ionosphere. We can also utilise more standard MIRIAD and CASA gain tables if required.

 The first step in obtaining the antenna gain compensation from an RTS solution is to realise that the measured Jones matrices can be decomposed into the direction independent gain ($\jones{D}{j}$) and a direction dependent component, common to all solutions. This direction dependent element is the beam model in the reference direction of the calibration solution. This direction is given in the sky model used for the calibration, it may be the position of a single source, or an assumed reference position for a group of components. The essential part is that it is a common component of all solutions, in the terminology of the calibration system it is the {\em alignment} matrix, ($\jones{A}{\mathrm{a}}$).

As the calibration is in the assumed alignment direction, rather than the desired pointing direction, we have to rotate the solution incorporating the direction dependence of the antenna. We first remove the alignment component: 
\begin{align}
\jones{J}{j,a} &= \jones{D}{j}\jones{A}{\mathrm{a}} \\
\jones{D}{j} &=\jones{J}{j,a} \jones{A}{\mathrm{a}}^{-1}. 
\end{align}

\noindent We can then form the Jones matrix in the desired direction of source, $s$: which we denote $\jones{J}{j,\mathrm{s}}$. This is obtained by post-multiplying by the model beam response for the source direction, $\jones{B}{\mathrm{s}}$.

\begin{equation}
\jones{J}{j,\mathrm{s}} =\jones{D}{j} \jones{B}{\mathrm{s}}.
\end{equation}

The model for the antenna used for the observations in this paper is an {\em analytic} beam model. The MWA antenna beam is modelled as the sum of 16 dual--polarisation dipoles above an infinite ground screen.  Alternative MWA beam models (e.g. Sokolowski et al. 2017)\nocite{sok:2017}  could also be utilised and a comparison of the polarimetric performance of the different models will be presented in future work.

\subsubsection{Applying Delay and Gain Compensation}

We then apply the delay and gain compensation to each sample. The voltage vector for channel $n$ and antenna $j$, is generated in response to an incident electric field vector ($\vec{e}$), which  contains a noise contribution ($\vec{n}_{j}$). In general, the noise contribution completely dominates the incident electric field vector from the source. Note that the beam is being formed on a single time sample, from a single channel.

\noindent Firstly the delay compensation is applied to the measured voltage.

\begin{align}
\vec{v}^{\prime}_j &= \vectt{v_x \times \exp\{-i\phi_{{j},{n},{x}}\}}{v_y \times \exp\{-i\phi_{{j},{n},{y}}\}}
\end{align}

The measured voltage is the incident signal, modified by the instrument Jones matrix plus noise.
\begin{align}
\vec{v}^{\prime}_{j} &= \jones{J}{\mathrm{actual},j}\vec{e} +  \vec{n}_{j}
\end{align}

\noindent using the solution $\jones{J}{j}$ as an estimate of $\jones{J}{\mathrm{actual},j}$ we premultiply by its inverse:

\begin{align}
\jones{J}{j}^{-1}\vec{v}^{\prime}_{j} &= \jones{J}{j}^{-1}\jones{J}{\mathrm{actual},j}\vec{e} +  \jones{J}{j}^{-1}\vec{n}_{j}
\end{align}

We then accumulate this product across all $N_{A}$ antennas, assuming we are perfectly calibrated then this sum forms an estimate, $\vec{e}^{\prime}$, of the true incident electric field $\vec{e}$, corrupted by noise $\vec{$\pmb{\sigma}$}$.

\begin{align}
\sum^{N_{A}}_{j} \jones{J}{j}^{-1}\vec{v}^{\prime}_{j} &= \sum^{N_{A}}_{j} (\vec{e} +  \jones{J}{j}^{-1}\vec{n}_{j}) \\
\vec{e}^{\prime} &= \vec{e} +  \vec{$\pmb{\sigma}$}
\end{align}
\noindent where
\begin{align}
\vec{$\pmb{\sigma}$} &= \frac{1}{N_{A}} \sum^{N_{A}}_{j} \jones{J}{j}^{-1}\vec{n}_{j}
\end{align}

\subsubsection{Forming the Stokes Parameters}

As shown in Equation  \ref{eq:IQUV} the Stokes parameters are formed from the coherency matrix  $\langle\vec{e}\vec{e}^{H}\rangle$. In the presence of noise, a single time sample of the coherency matrix becomes:

\begin{align}
\vec{e}^{\prime}\vec{e}^{\prime H} &=  \left(\vec{e} + \vec{$\pmb{\sigma}$}\right) \left(\vec{e} +  \vec{$\pmb{\sigma}$}\right)^{H}, \\
\vec{e}^{\prime}\vec{e}^{\prime H} &= \vec{e}\vec{e}^{H} + \bf{O}(\vec{$\pmb{\sigma}$}) + \vec{$\pmb{\sigma}$}\vec{$\pmb{\sigma}$}^{H} \label{eq:noise1}.
\end{align}

The expectation value,  $\langle\cdot\rangle$, or long running time average of this term will be dominated by the signal ($ \vec{e}\vec{e}^{H} $) and the noise squared term ($\vec{$\pmb{\sigma}$}\vec{$\pmb{\sigma}$}^{H}$) as the zero mean nature of the thermal noise will reduce terms linear in $\pmb{\sigma}$ to average to zero. If we consider the form of the noise squared term \vec{$\pmb{\sigma}$}\vec{$\pmb{\sigma}^{H}$}:

\begin{align}
\vec{$\pmb{\sigma}$}\vec{$\pmb{\sigma}$}^{H} &=  \frac{1}{N_{A}^{2}} \left(\sum^{N_{A}}_{j} \jones{J}{j}^{-1}\vec{n}_{j}\right) \left(\sum^{N_{A}}_{j} \jones{J}{j}^{-1}\vec{n}_{j}\right)^{H} \\
&=  \frac{1}{N_{A}^{2}}\left[\sum^{N_{A}}_{j} (\jones{J}{j}^{-1}\vec{n}_{j}) (\jones{J}{j}^{-1}\vec{n}_{j})^{H}  + \sum^{N_{A}}_{j}\sum^{N_{A}}_{k, j \neq k}(\jones{J}{j}^{-1}\vec{n}_{j})(\jones{J}{k}^{-1}\vec{n}_{k})^{H}\right].
\label{eq:noise2}
\end{align}

\noindent The first term in the square brackets on the RHS of Equation \ref{eq:noise2} is the sum of the autocorrelation of the noise contribution to the voltage in each antenna and the second term is the sum of all the correlated noise. If we assume that the receiver noise is uncorrelated then the second term averages down with time. As the antennas are noise dominated we can approximate this by the antenna autocorrelation. The autocorrelations contain the sky signal too, this is the {\em incoherent beam} after all. However, for a large number of antennas the benefit of removing the autocorrelation noise compensates for the loss of signal. This process also has the added benefit that the tied-array beam more clearly matches the single pixel response of the MWA interferometer.

To clarify the equivalence of the tied-array beam and interferometer response when we form beams in this manner consider the following thought experiment. Take a simple two element array, which is phased at zenith, with no calibration errors and equal gains, so no phase terms are required. The phased array sum, $\vec{V}$, of a single polarisation, $\vec{v}$, from each antenna including noise $\vec{n}$ is given by:
\begin{align}
\vec{V} = \left(\vec{v}_{1} + \vec{n}_1 + \vec{v}_{2} +  \vec{n}_{2}\right),
\end{align}
recall this voltage sample is detected by the action of $\vec{V}\vec{V}^{H}$. The product of which is the sum of the following sixteen terms:
\begin{align}
\vec{V}\vec{V}^{H} = \,& \vec{v}_{1}\vec{v}_{1}^{H} + \vec{v}_1\vec{v}_{2}^{H} + \vec{v}_1\vec{n}_{1}^{H} + \vec{v}_{1}\vec{n}_{2}^{H} + \nonumber \\
				&  \vec{v}_{2}\vec{v}_{1}^{H} + \vec{v}_2\vec{v}_{2}^{H} + \vec{v}_2\vec{n}_{1}^{H} + \vec{v}_{2}\vec{n}_{2}^{H} + \nonumber \\
				&  \vec{n}_{1}\vec{v}_{1}^{H} + \vec{n}_1\vec{v}_{2}^{H} + \vec{n}_1\vec{n}_{1}^{H} + \vec{n}_{1}\vec{n}_{2}^{H} + \nonumber \\
				&  \vec{n}_{2}\vec{v}_{1}^{H} + \vec{n}_2\vec{v}_{2}^{H} + \vec{n}_2\vec{n}_{1}^{H} + \vec{n}_{2}\vec{n}_{2}^{H}. 
\end{align}
Lets suppose we had access to a complex correlator and instead formed the complex correlation between the two inputs:
\begin{align}
\vec{C} =  & \left(\vec{v}_{1} + \vec{n}_1\right) \left(\vec{v}_{2} +  \vec{n}_{2}\right)^{H}, \nonumber \\
		& \vec{v}_{1} \vec{v}_{2}^H  + \vec{v}_1\vec{n}_2^{H} + \vec{n}_{1}\vec{v}_{2}^{H} + \vec{n}_1\vec{n}_2^{H}.
\end{align}
The four terms of $\vec{C}$ are present in the 16 terms of the tied-array sum, but there are others. Now another four terms can be accounted for by the conjugate of \vec{C}. Interferometers typically only form half the correlations, but the tied array does not have that luxury and  the sum contains both. This information is redundant and does not improve the signal to noise ratio. But that still leaves eight terms that are present in a tied-array sum, but not in a typical interferometric scheme. 
Consider the auto-correlations of the inputs to the tied array:
\begin{align}
\vec{A}_{11} = & \left(\vec{v}_1 + \vec{n}_1\right)\left(\vec{v}_1 + \vec{n}_1\right)^{H} \nonumber \\
		     = & \vec{v}_1\vec{v}_1^H + \vec{n}_1\vec{v}_1^H + \vec{n}_1\vec{v}_1^H + \vec{n}_1\vec{n}_1^H \\
\vec{A}_{22} = &  \vec{v}_2\vec{v}_2^H + \vec{n}_2\vec{v}_2^H + \vec{n}_2\vec{v}_2^H + \vec{n}_2\vec{n}_2^H	       
\end{align}
These eight terms are precisely those, that when subtracted from the tied-array sum produce the equivalent to the interferometer output. That is:
\begin{align}
\vec{C} \equiv \vec{V}\vec{V}^H - \vec{A}_{11} - \vec{A}_{22}
\end{align}

Therefore as we accumulate the individual voltages into the coherent $\vec{e}^{\prime}$, we also accumulate the individual antenna detected signals $\vec{e}^{\prime}_{j}\vec{e}^{ \prime H}_{j}$ for all $j$, we use these as an estimate of the autocorrelation component of the noise in the beam in the formation of the Stokes parameters. Removing the autocorrelations does not of course eliminate the noise, but does remove a substantial fraction. The remaining noise components are the terms linear in the noise in Equation \ref{eq:noise1} and the correlated noise in \ref{eq:noise2} which  are reduced by integration in time (and frequency).
		       
The individual Stokes parameters are formed in the following manner:

\begin{eqnarray*}
\mathrm{I} &=& [e^{\prime}_{x}e^{\prime *}_{x} - \frac{1}{N_{A}^2}\sum^{N_{A}}_{j}e^{\prime}_{j,x}e^{\prime *}_{j,x} ] 
\end{eqnarray*}
\begin{eqnarray}
& + &  [e^{\prime}_{y}e^{\prime *}_{y} - \frac{1}{N_{A}^2}\sum^{N_{A}}_{j}e^{\prime}_{j,y}e^{\prime *}_{j,y}] 
\end{eqnarray}
\begin{eqnarray*}
\mathrm{Q} &=& [e^{\prime}_{x}e^{\prime *}_{x} - \frac{1}{N_{A}^2}\sum^{N_{A}}_{j}e^{\prime}_{j,x}e^{\prime *}_{j,x} ]
\end{eqnarray*}
\begin{eqnarray}
& - &  [e^{\prime}_{y}e^{\prime *}_{y} - \frac{1}{N_{A}^2}\sum^{N_{A}}_{j}e^{\prime}_{j,y}e^{\prime *}_{j,y}] 
\end{eqnarray}
\begin{eqnarray} 
\mathrm{U} &=& 2 \times \mathrm{Re}[e^{\prime}_{x}e^{\prime *}_{y} - \frac{1}{N_{A}^2}\sum^{N_{A}}_{j}e^{\prime}_{j,x}e^{\prime *}_{j,y} ]
\end{eqnarray}
\begin{eqnarray} 
\mathrm{V} &=& -2 \times \mathrm{Im}[e^{\prime}_{x}e^{\prime *}_{y} - \frac{1}{N_{A}^2}\sum^{N_{A}}_{j}e^{\prime}_{j,x}e^{\prime *}_{j,y} ]
\end{eqnarray}

\begin{figure}[!h]
\centering
\begin{subfigure}[t]{0.5\textwidth}
\centering
\includegraphics[width=3in ,angle=270,left]{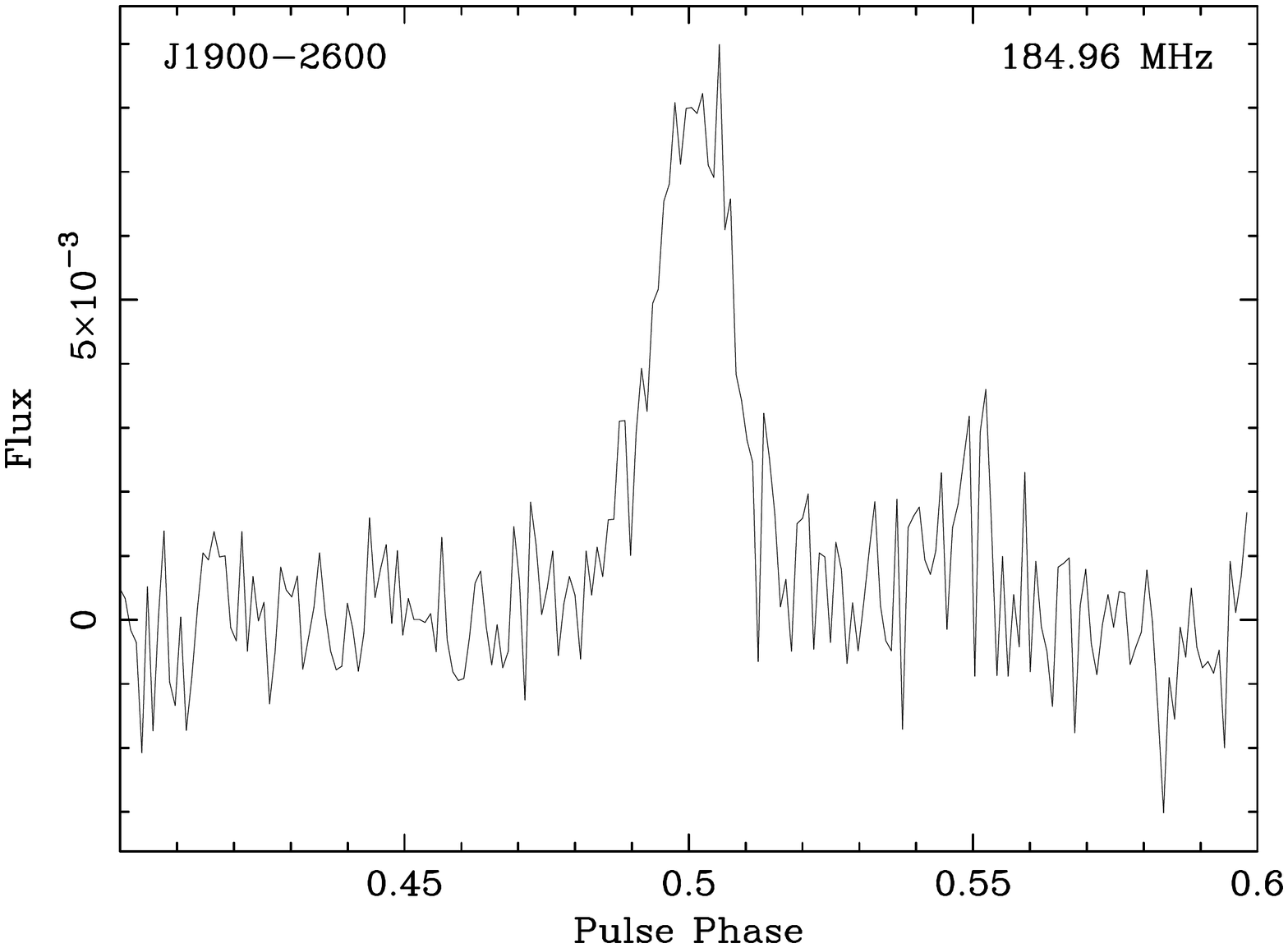} 
\end{subfigure}
\begin{subfigure}[t]{0.5\textwidth}
\centering
\includegraphics[width=3in, angle=270,left]{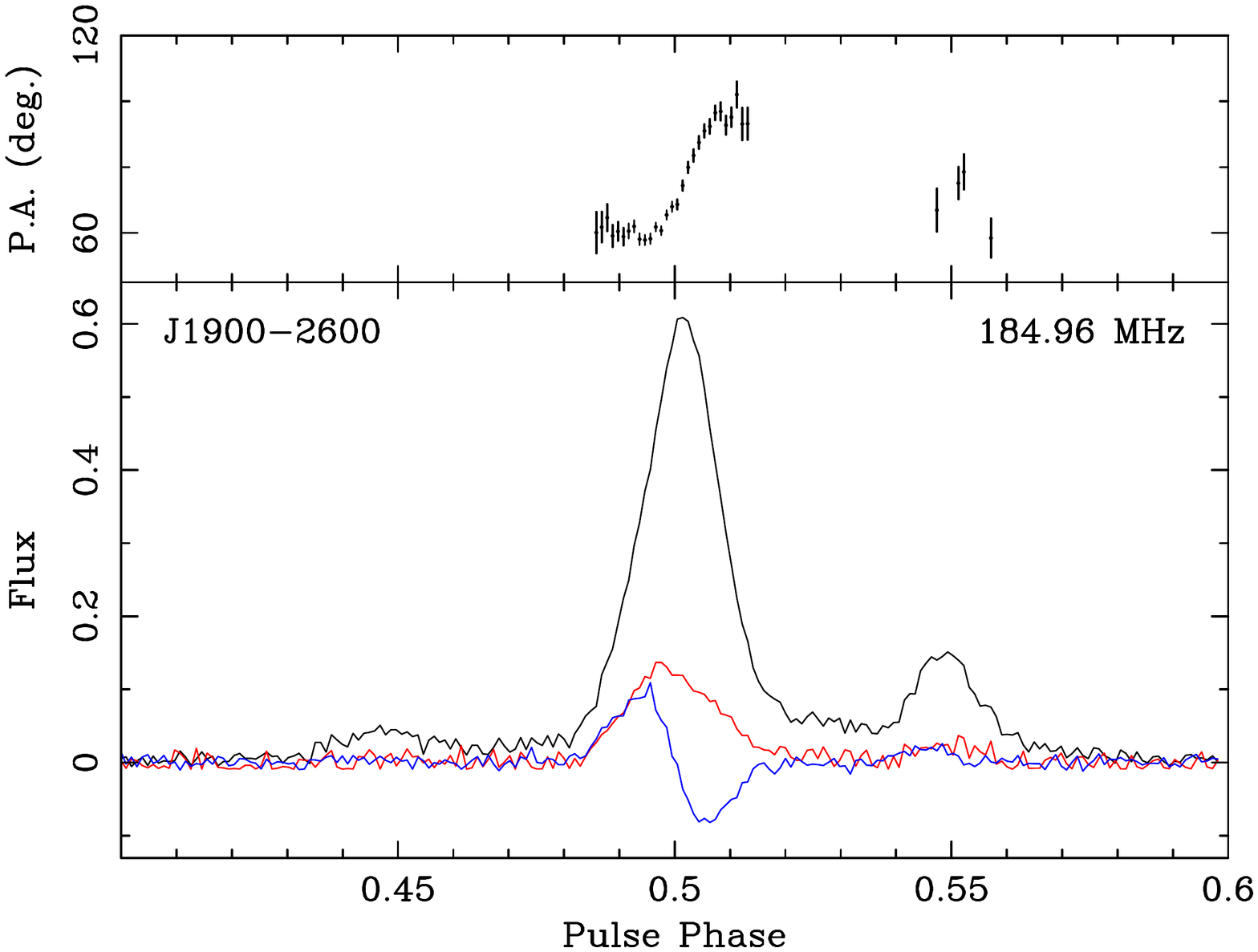} 
\end{subfigure}
\caption{The incoherent (upper) and coherent (lower) beamformed polarimetric profile of PSR J1900-2600, integrated for  approximately 1500 seconds using 30.72\,MHz of bandwidth. This S/N of approximately 240 and the incoherent sum S/N is 14; indicating that the sky temperature of this pointing considerably exceeds the receiver temperature at this frequency. The polarimetric profile is consistent with that published in Johnston et al (2008).  The black line is total intensity (Stokes I), the red line is linear polarised intensity and the blue line is circular polarised intensity (Stokes V). In the coherent profile the upper panel is the position angle of the linear polarisation.
  }\label{fig:2}
\end{figure}

\begin{figure}[!h]
\begin{subfigure}[t]{0.5\textwidth}
\includegraphics[width=3in ,angle=270, left]{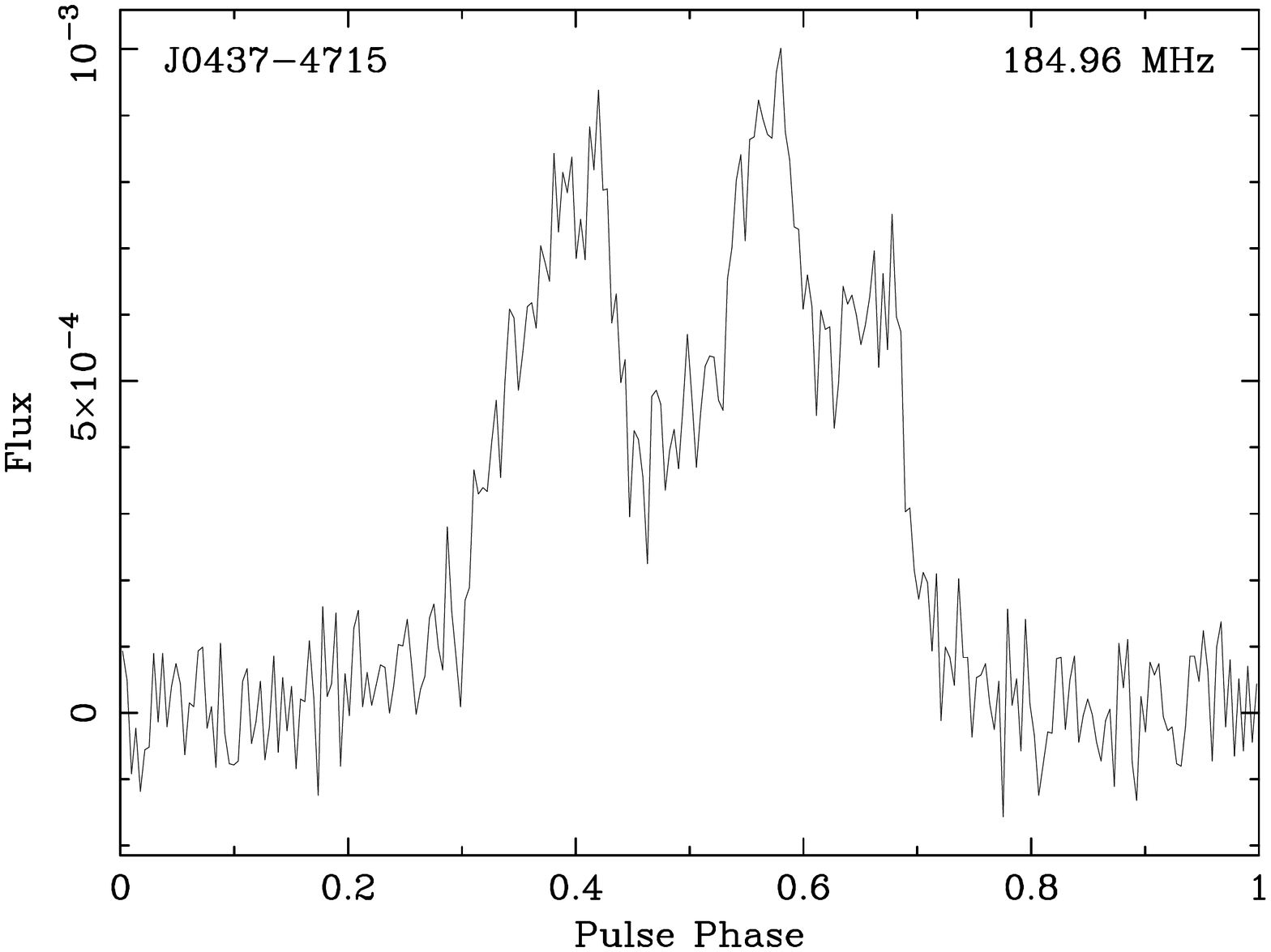} 
\noindent 
\end{subfigure} 
\begin{subfigure}[t]{0.5\textwidth}
\includegraphics[width=3in, angle=270, left]{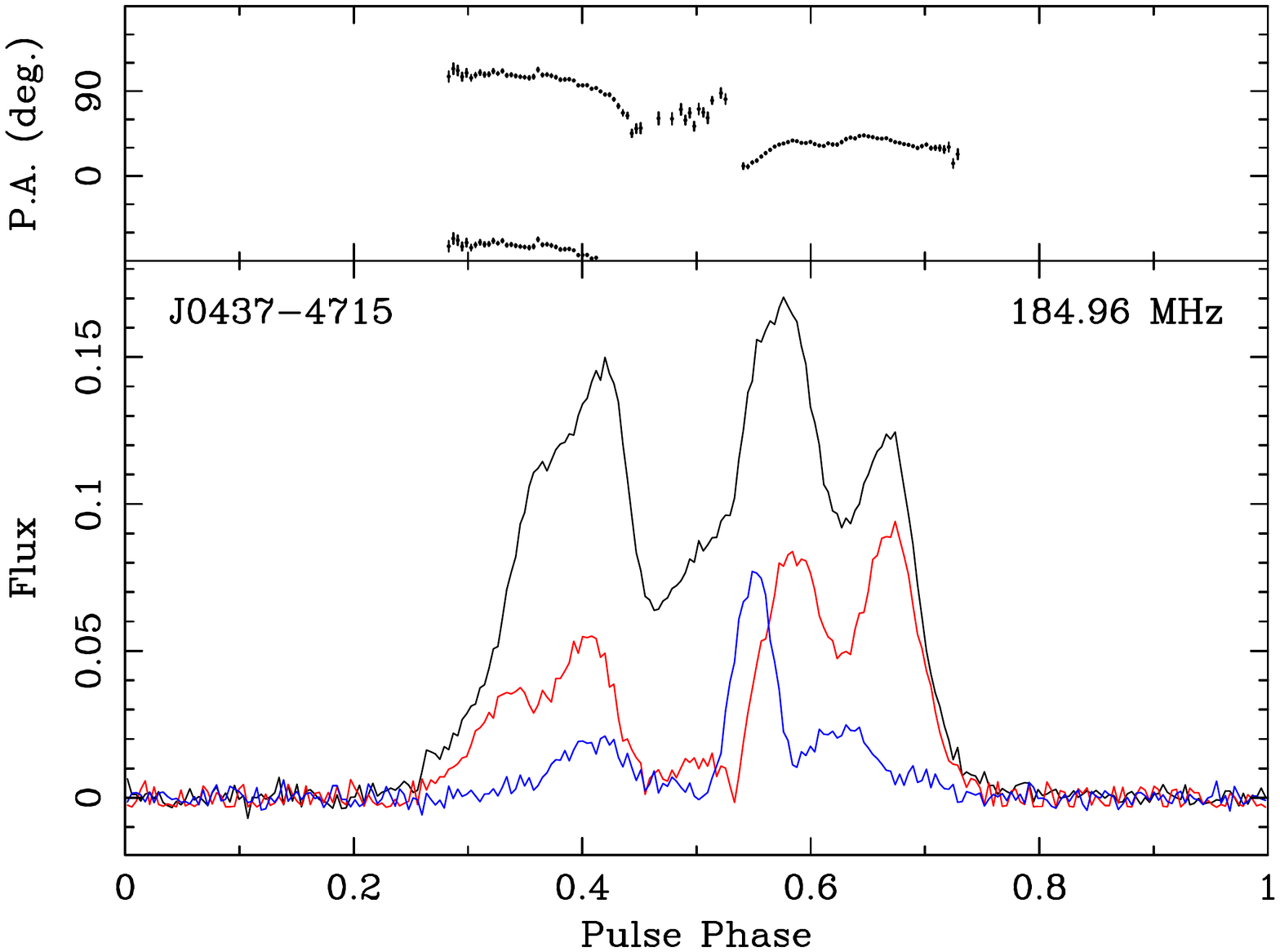} 
\end{subfigure}
\caption{The comparison between the incoherent (upper) and coherent beamformed (lower) profile of J0437-4715.  There is no polarisation information on this pulsar below 400\,MHz so direct comparison with other work is not possible, however the polarimetric profile published by Dai et al (2015) is in general agreement. The S/N improvement is consistent with that expected by coherent addition of the antenna voltages. The black line is total intensity (Stokes I), the red line is linear polarised intensity and the blue line is circular polarised intensity (Stokes V). In the coherent profile the upper panel is the position angle of the linear polarisation.} 
\label{fig:3}

\end{figure}

\section{Example Beamformed Observations}

The incoherent beamformer has already been used to study the properties and population of pulsars with the MWA (Xue et al 2017\nocite{xue:2017}). The coherent beamformer is also beginning to be employed (Bhat et al 2016; \nocite{bhat:2016}Meyers et al 2017\nocite{meyers:2017}; McSweeney et al 2017\nocite{mcsweeney:2017}; Bhat et al 2018\nocite{bhat:2018})  and is undergoing a process of validation. In this paper we present some example pulse profiles to demonstrate the capability of the beamformer; a companion publication, Xue et al (in preparation)  will present further validation. 

The beamformer generates either full-polarisation PSRFITS {\em search} mode files or VDIF format voltage beams. The profiles presented in this paper have been generated using the PSRFITS output pipeline. This format is a time series of detected powers in the Stokes basis,  all four Stokes parameters are formed and digitised to 8-bit, with a header conforming to the PSRFITS standard. These data were subsequently folded and dedispersed using the DSPSR software package (van Straten and Bailes 2011\nocite{vSB:2010}) and a timing ephemeris from the pulsar database\footnote{Ephemerides found using PSRCAT the pulsar catalogue (v. 1.57) retrieved from http://www.atnf.csiro.au/people/pulsar/psrcat/} (Manchester et al. 2005). Subsequent papers in this series will examine in detail the fidelity of the beamformed data. For the purposes of this paper, we will compare the morphology of the determined polarisation with that already in the literature.

\subsection{Coherent vs Incoherent Signal to Noise}
\label{sec:SNR}
The efficiency of incoherent beamforming, i.e. how the signal to noise ratio (S/N) of an observation increases by adding the detected power from the constituent antennas into the sum, is a function of the ratio of sky to antenna noise. This behaviour is summarised in Kudale and Chengular (2017)\nocite{kudale:2017}, who compare the sensitivities of coherent and incoherent summation for the Giant Metrewave Radio Telescope (GMRT).  They give the S/N obtained in an observation of a source of flux density $S$, using an incoherent sum of $N_{A}$ antennas, each with a receiver temperature $T_{R}$  and gain $G$, with an antenna temperature of $T_{A}$ to be:

\begin{align}
\mathrm{(S/N)_{incoh}} \propto \frac{\sqrt{N_{A}}GS}{[(T_{R} + T_{A})^{2} + (N_{A}-1)T_{A}^2]^{1/2}},
\end{align}
\noindent from which it can be seen that if the antenna temperature dominates the receiver temperature sufficiently, then the S/N is no longer a function of $\sqrt{N_{A}}$. Although the MWA is in general sky-dominated, the ratio is not high. However, there are pointing directions, near the Galactic plane for example, where this is not the case. 

In the case of a coherent sum, the S/N with perfect phasing is given by:

\begin{align}
\frac{GS + (N_{A}-1)(GS)}{[(T_{R} + T_{A})^{2} + (N_{A} - 1)(GS)(T_{A} + T_{R}) + (N_{A} - 1)^{2}(GS)^{2}]}
\end{align}

\noindent and when the source contribution to the system temperature is small ($GS << (T_{A} + T_{R})$) the S/N reduces to the expected expression:

\begin{align}
\mathrm{(S/N)_{coh}} \propto  \frac{N_{A}(GS)}{(T_{A} + T_{R})}
\label{eq:SNR}
\end{align}

\noindent Thus, in the case where receiver noise dominates over sky noise, the coherent sum is a factor of $\sqrt{N_{A}}$ better than an incoherent sum. However, in the case where sky noise dominates, the improvement may be as much as a factor of $N_{A}$. There is one more case to consider, when the source makes a contribution to the noise temperature. This is {\em self-noise} and is examined in detail in Kulkarni (1989)\nocite{kulkarni:1989} and Anantharamaiah et al (1991)\nocite{ananth:1991}. In a beamfomer, self-noise is an issue when the source flux is comparable to the System Equivalent Flux Density (SEFD)\footnote {SEFD is the flux of a source that, if observed, would double the system temperature.} of an antenna divided by the number of antennas in the array. The SEFD of an MWA antenna is approximately $\sim 20$\,kJy at 200\,MHz, so for self-noise to start to become an issue the source flux would need to be greater than $\sim 200$\,Jy, which is easily reached in observations of bright single pulses e.g. Meyers et al (2017, 2018 in preparation)\nocite{meyers:2017}, however in observations of pulsed sources the noise measurement is taken when the pulse is not present, thereby mitigating this effect. 

The efficiency of the coherent beamforming is also a function of the accuracy of the antenna phasing, which is determined by the quality and applicability of the calibration. As examined in detail by Kudale and Chengular (2017), if the phase error has zero mean and variance $\sigma^{2}$ then Equation \ref{eq:SNR} becomes:

\begin{align}
\mathrm{(S/N)_{coh}} \propto  \frac{N_{A}(GS) \exp{\{-\sigma^{2}/2}\}}{(T_{A} + T_{R})}.
\end{align}

\noindent Since the MWA is a low--frequency telescope, with comparatively short baselines, the variance of any phase error will be low. However in some calibration schemes a calibration solution is transferred from a dedicated calibration observation. When solutions are transferred the ionospheric refraction will be different and could potentially result in a persistent phase error, reducing the efficiency of the beamforming. It is becoming more common to use "in-field" calibration where the bulk refractive offset of the field is not an issue, however we do not employ a scheme to determine the accuracy of the calibration solutions obtained. 

\subsection{Pulsar Observations}

\subsubsection{PSR J1900--2600}

Figure \ref{fig:2} shows pulse profiles for PSR J1900--2600 obtained using an MWA-VCS observation at 184.5\,MHz with 30.72\,MHz of bandwidth (3072 $\times$ 10 kHz channels) and an integration time of approximately 1500 seconds. The profiles resulting from the incoherent and coherent sum are shown in the upper and lower panels, respectively.  The S/N\footnote{The pulsar S/N is calculated by the PSRCHIVE package, following the method outlined in Cordes \& McLaughlin (2003)\nocite{cordes:2003}} of the coherently-formed beam is approximately 240, the incoherent sum of the same observation is only 14. The factor of improvement is 17, much larger than the factor of 10 expected. J1900-2600 is very close to the Galactic plane therefore, as examined in \S \ref{sec:SNR}, the sky temperature likely dominates over the receiver noise, making the incoherent sum less efficient. 

This pulsar shows considerable profile evolution as a function of frequency. However, the profile presented here is consistent with that at 243\,MHz in Johnston et al (2008)\nocite{johnston:2008}.

\subsubsection{PSR J0437--4715}

PSR J0437--4715 is a very well studied pulsar in the southern hemisphere, and the nearest and brightest millisecond pulsar. However there is no published low--frequency polarisation data for this object. Figure \ref{fig:3} shows the first polarimetric profile of J0437--4715 below 200\,MHz.  In this case the improvement in S/N between the incoherent and coherent sum is approximately 8, which is comparable with that expected. As the pulsar is in a region of the sky well away from the Galactic plane so the incoherent sum S/N should not be sky dominated. The polarisation profile is consistent with observations made at higher frequencies with the Parkes telescope. Observations have been published at 430, 660 and 732\,MHz (Navarro et al. 1997\nocite{navarro:1997}; van Straten 2002\nocite{vanS:2002} and Dai et al 2015\nocite{dai:2015}). While the linear polarisation is in good agreement, especially with the lowest frequency observations, the circular polarisation seems to display a sense change which is not evident here. However, given the rapid evolution of the pulse profile as a function of frequency it is not surprising that a true comparison is difficult.

\section{Summary}

The primary motivation for this paper is to document the algorithms used in the MWA tied-array beamformer. We have described the steps required to calibrate and coherently combine voltages from the 128 antennas of the Murchison Widefield Array. We have also discussed some of the issues inherent in extrapolating signal--to--noise ratios between incoherent and coherent observations. Finally we have presented some representative profiles of low--frequency observations of PSRs J0437--4715 and J1900--2600, and compared them with their published profiles at nearby or higher frequencies. In a subsequent paper in this series we will examine the polarimetric stability and fidelity of the MWA beamformed observations. 

\begin{acknowledgements}

\noindent This scientific work makes use of the Murchison Radio-astronomy Observatory, operated by CSIRO. We acknowledge the Wajarri Yamatji people as the traditional owners of the Observatory site. Support for the operation of the MWA is provided by the Australian Government (NCRIS), under a contract to Curtin University administered by Astronomy Australia Limited. We acknowledge the Pawsey Supercomputing Centre which is supported by the Western Australian and Australian Governments. This research was conducted by the Australian Research Council Centre of Excellence for All-sky Astrophysics (CAASTRO), through project number CE110001020. FK acknowledges support through the Australian Research Council grant DP140104114

\end{acknowledgements}


\printbibliography

\end{document}